\documentclass[twocolumn,showpacs,preprintnumbers,amsmath,amssymb,prl]{revtex4}
\usepackage{graphicx}
\input epsf \unitlength=1mm

\begin{document}

\title{Theoretical evidence for a reentrant phase diagram in {\em
ortho-para} mixtures of solid H$_2$}

\author{Bal{\'a}zs Het{\'e}nyi$^{1,3}$\footnote{Present address:Institut
    f\"ur Theoretische Physik, Technische Universit\"at Graz, Petersgasse 16,
  A-8010, Graz, Austria}, Sandro Scandolo$^{2,3}$, and Erio
Tosatti$^{1,3}$}

\affiliation{$^1$SISSA-International School of Advanced Studies, via
Beirut 2-4, Trieste I-34014, Italy \\
        $^2$The Abdus Salam International Centre for Theoretical Physics, 
I-34014 Trieste, Italy \\
        $^3$Democritos-INFM, via Beirut 2-4, Trieste I-34014, Italy 
}

\begin{abstract}
  We develop a multi order parameter mean-field formalism for systems of
  coupled quantum rotors.  The scheme is developed to account for systems
  where {\it ortho-para} distinction is valid.  We apply our formalism to
  solid H$_2$ and D$_2$.  We find an anomalous {\it reentrant} orientational
  phase transition for both systems at thermal equilibrium.  The correlation
  functions of the order parameter indicate short-range order at low
  temperatures.  As temperature is increased the correlation increases along
  the phase boundary.  We also find that even extremely small {\it odd-J}
  concentrations (1\%) can trigger short-range orientational ordering. 
\end{abstract}
\maketitle
\begin{figure}
\setlength\fboxsep{0pt} \centering
\scalebox{0.40}{\includegraphics[angle=0]{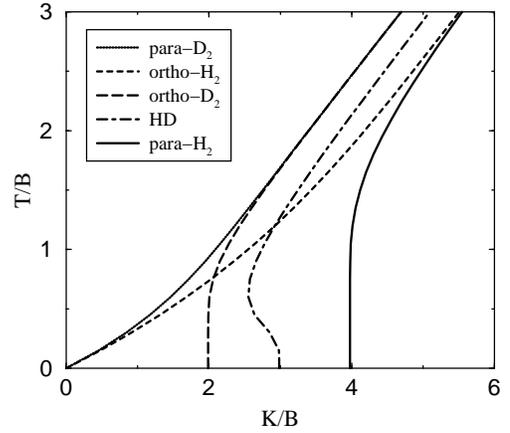}}
\caption{Mean-field phase diagrams for coupled quadrupolar rotor
models corresponding to solid molecular hydrogen and various isotopes.} 
\label{fig:mf}
\end{figure}

\label{sec:intro}
Quantum effects dominate the low temperature ($T<200$K) phase diagram of
solid molecular hydrogen in a wide range of pressures from ambient up to
$\sim100$ GPa~\cite{Silvera80,Mao94}.  In this regime the coupling between
molecules is smaller than the molecular rotational constant, so quantum
effects are generally described by means of weakly coupled quantum-rotor
models~\cite{VanKranendonk83}.  Homonuclear molecules (H$_2$ and D$_2$) can
assume only even or odd values of the rotational quantum number $J$,
depending on the parity of the nuclear spin.  Important differences exist in
the phase diagrams of even-$J$ ({\em para}-H$_2$ and {\em ortho}-D$_2$),
odd-$J$ ({\em ortho}-H$_2$ and {\em para}-D$_2$) and all-$J$ (HD) species.
(For a summary of experimental results on the orientational ordering in
H$_2$, D$_2$, and HD see Fig. 1b of reference \onlinecite{Cui95}.)  At low
pressure or high temperature, even-$J$ species are found in a rotationally
disordered free-rotor state (phase I). Increasing pressure causes an increase
of the intermolecular coupling, and eventually leads to an orientationally
ordered state (phase II).  Odd-$J$ systems on the other hand are
orientationally ordered at low temperature and ambient pressure and remain
ordered as pressure is increased.  The stronger tendency of {\em ortho}-H$_2$
to order can be traced to the fact that its $J=1$ lowest rotational state
allows for a spherically asymmetric ground state, unlike the $J=0$ ground
state of even-$J$ species. The pressure-temperature phase diagram of HD
exhibits a peculiar {\em reentrant} shape~\cite{Moshary93}.  Reentrance
refers to phase diagrams where in some range of pressure the system reenters
the disordered phase at ultra-low temperatures (see HD in Fig. \ref{fig:mf}).
The zero-temperature orientationally disordered phase is characterized by an
energy gap against $J=1$ excitations.  When this gap is sufficiently small
and the temperature is finite, the thermally generated $J=1$ excitations
suffice to induce ordering, which is then reentrant, as also shown by
mean-field theory~\cite{Freiman91,Brodyanskii93,Freiman98}.  Reentrance is
also found in models of two-dimensional rotors~\cite{Simanek85}, such as the
quantum anisotropic planar rotor (QAPR)
model~\cite{Martonak97,Muser98,Hetenyi99}.

High pressure experiments have focused mainly on the behavior of pure
species.  While understanding the orientational transition in {\em
  ortho}-{\em para} mixtures of quantum rotors in two and three dimensions is
a problem of long-standing theoretical and experimental
interest~\cite{Feldman95,Harris85,Sullivan87,Goncharov96,Hochli90,Binder02,Kokshenev96a,Mazin97,Goncharov01},
less attention has been given at establishing the effects of {\em ortho-para}
distinction.  Exceptions~\cite{Feldman95,Goncharov96,Mazin97,Goncharov01} are
the investigation of vibrons in {\em ortho-para} mixtures by Feldman {\em et
  al.}~\cite{Feldman95} and a recent work by Goncharov {\em et
  al.}~\cite{Goncharov96}, where {\em ortho}-D$_2$ mixed with small amounts
of {\em para}-D$_2$ indicated that the possibility of an orientationally
frustrated phase between phases I and II (phase II').  Phase II' persisted
for a narrow pressure range ($\sim 2$GPa) for a thermally equilibrated {\em
  ortho-para} mixture.  

At ambient pressure hydrogen is known to enter gradually a phase of
short-range order (orientational glass)~\cite{Hochli90,Binder02} when the concentration of {\em ortho}
species is below $\sim 53$ \% ~\cite{Harris85,Sullivan87,Hochli90}. 
Short-range order is believed to be a manifestation of the orientational
frustration due to the random location of {\em ortho} molecules in the
lattice, and eventually transforms into long-range order at {\em ortho}
concentrations approaching unity.  The pressure-induced increase of the
intermolecular anisotropic coupling causes a decrease of the minimum
concentration of {\em ortho} species required to trigger the orientational
transition~\cite{Silvera80}.  At extremely high pressures ($\sim110$ GPa)
orientational freezing is observed also in pure {\em para}-H$_2$, as stated
above.  Ambient pressure experiments on pure species as well as on samples
with fixed concentration are possible because the rate of interconversion
between {\em para} and {\em ortho} H$_2$ species is extremely slow at low
pressure. Samples with fixed {\em ortho}-{\em para} concentration are
prepared by letting the system thermalize at the corresponding temperature
for very long times and experiments are then performed at any other
temperature by rapid heating or quench ( {\em rapid} on the time scale of
interconversion).  At high pressures however the interconversion rate rises
steeply~\cite{Strzhemechny00}.  At $58$ GPa, the highest pressures where the
conversion rate has been measured, interconversion takes place in the time
scale of seconds.  While interconversion is still orders of magnitude slower
than molecular rotation, the {\em ortho-para} distribution in this range of
pressure is likely to be determined by thermal equilibrium.  On the other
hand {\it ortho-para} distinction is still valid. 

\begin{figure}
\setlength\fboxsep{0pt} \centering
\scalebox{0.40}{\includegraphics[angle=0]{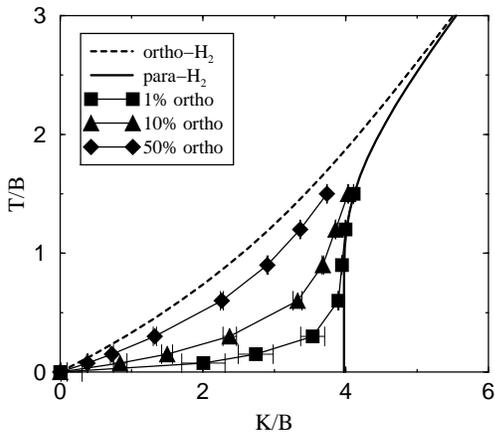}}
\caption{Multi order parameter (MOP) mean-field phase diagrams for coupled
quadrupolar rotor models corresponding to solid molecular hydrogen at
various {\it ortho} concentrations.} 
\label{fig:mpmf}
\end{figure}

Previous theoretical studies addressing quantum effects in compressed
hydrogen (mean-field~\cite{Freiman91,Brodyanskii93,Freiman98} or quantum
Monte Carlo~\cite{Kaxiras94,Cui97,Runge92,Pollock94} studies) have in most
cases neglected the distinction between even-$J$ and odd-$J$
species~\cite{Kaxiras94}, and in those cases where the distinction was
explicitly considered, the analysis was restricted to pure even-$J$ or
odd-$J$ species~\cite{Cui97,Runge92,Pollock94}.  On the other hand,
theoretical modeling of mixed {\em ortho}-{\em para}
systems~\cite{Harris85,Kokshenev96a,Feldman95} at low pressure is generally
limited to $J=0$/$J=1$ states.~\cite{Harris85,Kokshenev96a} This is not an
accurate model for compressed H$_2$, as pressure and temperature cause a
non-negligible admixture of rotational states with higher $J$'s in the ground
state of both {\em ortho} and {\em para} species.  A binary alloy
model~\cite{Feldman95} was also used to analyze the low-pressure vibrons in
{\it ortho-para} mixtures. 

In order to calculate the phase diagram of {\em ortho-para} mixtures of
hydrogen in the solid phase we develop a multi order parameter (MOP)
mean-field theory which accounts explicitly for {\it ortho-para} distinction
in systems of coupled rotors.  Our formalism enables us to treat a system of
$\sim 3000$ molecules, which would be extremely computationally intensive
with quantum Monte Carlo.  Our formalism also includes correlations due to
{\it ortho-para} distinction, a feature that is absent in standard mean-field
theory.  We apply the formalism to a system of coupled quadrupolar rotors
whose centers of mass form a face-centered cubic lattice.  We find that the
phase line separating the orientationally disordered state (phase-I) from the
orientationally ordered state is {\em reentrant} in the case of thermally
equilibrated {\it ortho-para} mixtures for both H$_2$ and D$_2$. We also find
that orientational order is short-ranged at low temperatures.  The validity
of our conclusion is supported by the fact that for the two-dimensional model
mean-field theory~\cite{Simanek85,Martonak97} is in agreement with quantum
Monte Carlo~\cite{Muser98,Hetenyi99} simulations on the overall shape of the
phase diagram.

The Hamiltonian of a system of $N$ coupled quadrupolar quantum rotors
interacting through a quadrupole-quadrupole potential can be written
in the form
\begin{eqnarray}
H &=& B \sum_{i=1}^N \hat{L_i}^2 + \frac{K}{2} \sum_{i<j}^N
\left( \frac{R_0}{R_{ij}} \right)^5 \\
 & & \sum_{m,n} C(224;mn) Y_{2m} (\Omega_i) Y_{2n}(\Omega_j) 
Y_{4m+n}^*(\Omega_{ij}),  \nonumber
\label{eqn:H}
\end{eqnarray}
where $B$ is the molecular rotational constant, $K$ is the coupling strength,
$C(224;mn)$ are Clebsch-Gordan coefficients, $\Omega_i$ denote the
coordinates of rotor $i$, and $\Omega_{ij}$ denote the direction of the
vector connecting rotors $i$ and $j$.  We define $R_0$ to be the nearest
neighbor distance, $R_{ij}$ is the distance between molecules $i$ and $j$. 
Our MOP mean-field theory is based on the trial Hamiltonian
\begin{eqnarray}
H_0 &=& B \sum_{i=1}^N \hat{L_i}^2 + K \sum_{i<j}^N \left(
\frac{R_0}{R_{ij}} \right)^5 \\ & & C(224;00) Y_{20}
(\Omega_i) \gamma_j Y_{40}^*(\Omega_{ij}), \nonumber
\label{eqn:H0}
\end{eqnarray}
where $\gamma_i$ are parameters.  Variation of the free-energy leads to the
self-consistent expression
\begin{equation}
\gamma_i = \langle Y_{20} (\Omega_i) \rangle_0. 
\label{eqn:gamma}
\end{equation}
Since the trial Hamiltonian in Eq. (\ref{eqn:H0}) is a sum of
single-rotor Hamiltonians, it follows that for each $\gamma_i$ in
Eq. (\ref{eqn:gamma}) the average needs to be performed over the
corresponding coordinate $\Omega_i$ only.  {\it Ortho-para}
distinction can be implemented by restricting a particular average to
be over odd-$J$ or even-$J$ states. 

Since in this study we are only interested in the phase diagram we
expand to first-order in the potential~\cite{Simanek85}.  Such an
expansion is expected to be valid here, since the transition between
the long-range ordered state and the disordered state is only weakly
first-order~\cite{Cui95}.  Furthermore the transition between the
short-range ordered state and the disordered state is known to be
continuous at ambient pressure~\cite{Harris85}.  The resulting
expression can be written
\begin{equation}
\gamma_i = K 3 \sqrt{\frac{2}{35}} \Phi_i \sum_j \gamma_j \left(
\frac{R_0}{R_{ij}} \right)^5 Y_{40}(\Omega_{ij}),
\label{eqn:mtx}
\end{equation}
where $\Phi_i$ is the phase correlator defined as $\Phi_i =
\int_0^{\beta} d \tau \langle Y_20(\Omega(0)) Y_{20} (\Omega(\tau))
\rangle_{free,i}$.  The averaging is to be performed over a free rotor
at inverse temperature $\beta$, and {\it ortho-para} distinction can
be invoked by restricting the average as described above.  Thus, for a
given temperature and rotational constant, $\Phi_i$ can take on two
values depending on whether rotor $i$ is odd-$J$ or even-$J$.  Order
will be signalled by non-zero solutions for $\gamma_i$, disorder by
solutions in which all $\gamma_i$ are identically zero.  The standard
mean-field theory~\cite{Freiman91,Brodyanskii93} is automatically
recovered for pure systems. 

\begin{figure}
\setlength\fboxsep{0pt} \centering
\scalebox{0.40}{\includegraphics[angle=0]{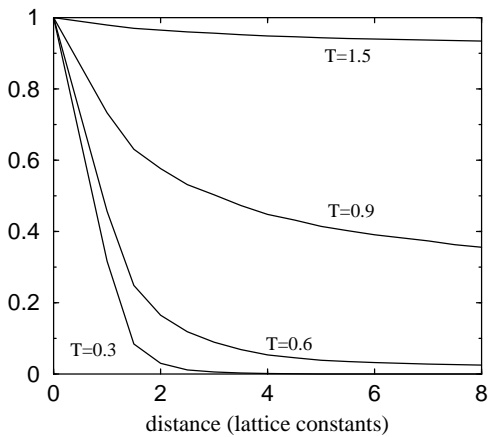}}
\caption{Correlation functions along the phase boundary of the 10\%
{\it ortho} system of solid H$_2$.} 
\label{fig:cf}
\end{figure}

We calculated the phase diagrams for systems of fixed odd-$J$ fraction, as
well as at the thermal equilibrium distribution, taking account of nuclear
spin degeneracy.  D$_2$ and H$_2$ differ by the values of the rotational
constants ($2B_{D_2} = B_{H_2}$) and in the degeneracies of states. {\it
  Ortho}({\it para}) H$_2$ is restricted to be odd-$J$(even-$J$) angular
momentum and in D$_2$ the reverse.  In our calculations for mixed systems,
configurations of 2916 molecules on a face-centered cubic lattice were used. 
For a given odd-$J$ fraction 20 configurations are generated.  A cutoff
distance of 2.65 times the lattice constant is used, thus 134 nearest
neighbors are included in the interaction calculation.  The phase transition
point is obtained by diagonalizing Eq. (\ref{eqn:mtx}) (which is a matrix
equation) and finding the lowest coupling constant for which a nonzero
solution exists. 

\label{sec:results}
In Fig. \ref{fig:mf} we present the results of standard mean-field
theory phase diagrams of pure odd-$J$/even-$J$ of H$_2$ and D$_2$, and
HD, all in excellent qualitative agreement with the experimental
results (see Fig. 1b of reference \onlinecite{Cui95}).  The energy
scale is defined to be the rotational constant of the H$_2$ molecule. 
The main difference between the phase diagrams of odd-$J$ and even-$J$
systems is accurately captured, namely that at low pressures odd-$J$
systems are always ordered, whereas even-$J$ systems order at finite
pressures.  As expected from experiment odd-$J$ D$_2$ orders at a
lower coupling strength than odd-$J$ H$_2$, and the reentrant phase
transition in HD is also well reproduced by mean-field
theory~\cite{Freiman91,Brodyanskii93,Freiman98}. 

\begin{figure}
\setlength\fboxsep{0pt} \centering
\scalebox{0.40}{\includegraphics[angle=0]{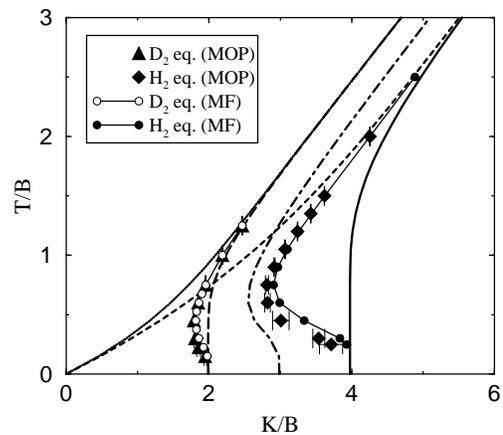}}
\caption{Phase diagrams of Fig. \ref{fig:mf} and those of the pure
systems H$_2$ and D$_2$ at thermal equilibrium distribution calculated
using our multi order parameter (MOP) mean-field theory and the
standard mean-field theory (MF).  For the mean-field phase diagrams
see legend of Fig. \ref{fig:mf}.} 
\label{fig:rnt}
\end{figure}

In Fig. \ref{fig:mpmf} the results of the formalism presented above are shown
for solid H$_2$.  As the {\it ortho} concentration is decreased the system
tends towards disorder entering the ordered state at higher coupling
constants for a given temperature.  A noteworthy result of our calculations
is that even at an {\it ortho} concentration of 1\% the system enters an
ordered state at coupling constants quite different from that of pure {\it
  para} hydrogen, and that for any {\it ortho} concentration the ground state
is always ordered.  The 50\% {\it ortho} system is very close to the pure
{\it ortho} one.  Our results are consistent with the phase diagram shown in
Fig. 1 of Ref. \onlinecite{Mazin97} in which the pure {\em para}-H$_2$ phase
transition occurs at $\sim110$ GPa, and {\em ortho} containing samples show
ordering at lower pressures. 

\begin{figure}
\setlength\fboxsep{0pt} \centering
\scalebox{0.40}{\includegraphics[angle=0]{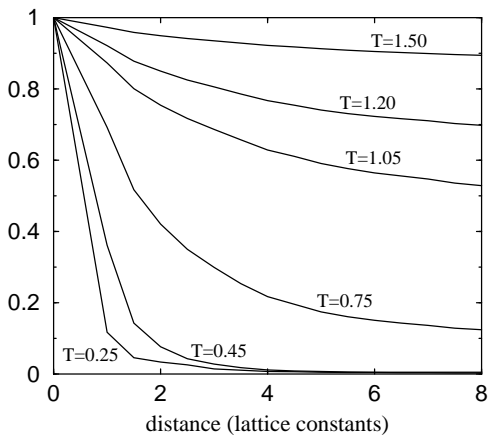}}
\caption{Correlation functions along the phase boundary of the
of solid H$_2$ at thermal equlibrium.} 
\label{fig:cfrnt}
\end{figure}

In order to assess the nature of the ordering we calculated
correlation functions of the local order parameters $\{\gamma_i\}$. 
This is an advantage of our multi order parameter formalism over the
standard mean-field theory, since in single-site mean-field theory
only phases of complete order or disorder are possible.  In particular
we have calculated
\begin{equation}
G(r) = \langle (\gamma(0) \gamma(r))^2 \rangle. 
\end{equation}
The results along the phase line of the 10\% {\it ortho} system of
hydrogen are shown in Fig. \ref{fig:cf}.  The correlation increases
along the phase boundary with increasing temperature indicating the
onset of long-range order.  The onset of long range order is due to
the fact that {\it ortho-para} distinction ceases as temperature and
coupling constant (pressure) are increased. 

The orientational ordering of the system with {\it ortho-para} concentrations
at thermal equilibrium is shown in Fig. \ref{fig:rnt} along with the phase
diagrams of the pure systems (also shown in Fig. \ref{fig:mf}) for
comparison.  For H$_2$ and D$_2$ we calculated the phase diagrams via the
multi order parameter mean-field theory presented here, and via the standard
mean-field theory.  The resulting phase diagrams indicate reentrance in both
D$_2$ and H$_2$.  Reentrance is also seen~\cite{HetenyiUP} in the
corresponding 2D model~\cite{Martonak97}.  Reentrance is stronger in the case
of H$_2$, due to higher relative weight of odd-$J$ contribution (nuclear spin
degeneracy), and higher rotational constant (stronger quantum effects).  The
correlation functions for different temperatures along the reentrant phase
diagram are shown for H$_2$ in Fig. \ref{fig:cfrnt}.  As the temperature
increases correlation increases along the phase boundary.  At high
temperatures ($T>1$) the order is definitely long-range.  We conclude that
short-range order may be present up to $T=0.75$.  In the case of D$_2$
reentrance is less severe. 

\label{sec:conclusion}

In summary we calculated the phase diagrams of solid hydrogen and its
isotopes.  We found that the I-II phase line is interspersed by another
phase, likely to be orientationally frustrated, as suggested by the
experiments of Goncharov {\it et al.}  (Ref. \onlinecite{Goncharov96}).  For
a thermal distribution of {\it ortho-para} rotors we find a reentrant phase
diagram for both H$_2$ and D$_2$.  At low temperatures the order is
short-ranged.  Most experimental signatures of the long-range ordered phase
are well known,~\cite{Silvera80,Mao94} the short-range ordered phase may be
seen by probing the distribution of the local order parameter.  While nuclear
magnetic resonance~\cite{Harris85} is a useful probe, it may be difficult to
apply at high pressure.  The II` phase was found~\cite{Goncharov96} by
investigating the Raman vibronic shift and it is known to be sensitive in
{\em ortho-para} mixtures~\cite{Feldman95}.  Rotational Raman lines are also
sensitive to local order.  Experimental studies to resolve the issues raised
here would be helpful. 


Work at SISSA/ICTP/Democritos was sponsored by MIUR FIRB
RBAU071S8R004, FIRB RBAU01LX5H, and MIUR COFIN 2003, as well as by
INFM.  We acknowledge using resources of CINECA/HPC-Europa.  We are
grateful to Professor Y. A. Freiman for helpful discussions.



\begin{thebibliography}{10}

\bibitem{Silvera80} I. Silvera, {\it Rev. Mod. Phys.} {\bf 52} 393 (1980). 

\bibitem{Mao94} H.-K. Mao and R.~J. Hemley, {\it Rev. Mod. Phys.} {\bf 66}
  671 (1994). 

\bibitem{VanKranendonk83} J. Van Kranendonk {\it Solid Hydrogen: Theory of
    the Properties of Solid H$_2$, HD, and D$_2$} (Plenum Press, New York,
  1983). 

\bibitem{Cui95} L. Cui {\it et al.}, {\it Phys. Rev. B} {\bf 51} 14987
  (1995). 

\bibitem{Moshary93} F. Moshary {\it et al.}, {\it Phys. Rev. Lett.} {\bf 71 }
  3814, (1993). 

\bibitem{Freiman91} Y.~A. Freiman {\it et al.}, {\it J. Phys. Condens. Mat.} 
  {\bf 3 }, 3855 (1991). 

\bibitem{Brodyanskii93} A.~P. Brodyanskii {\it et al.}, {\it Low. Temp. 
    Phys.} {\bf 19 }, 368 (1993). 

\bibitem{Freiman98} Y.~A. Freiman {\it et al.}, {\it J. Low Temp. Phys.} 
  {\bf 113 }, 723 (1998). 

\bibitem{Simanek85} E. \v{S}im\'{a}nek, {\it Phys. Rev. B} {\bf 32}, 500
  (1985). 

\bibitem{Martonak97} R. Martonak {\it et al.}, {\it Phys. Rev. E } {\bf 55 },
  2184 (1997). 

\bibitem{Muser98} M.~H. M\"user and J. Ankerhold {\it Europhys. Lett. } {\bf
    44 } 216, (1998). 

\bibitem{Hetenyi99} B. Het\'enyi {\it et al.} {\it Phys. Rev. Lett. } {\bf 83
  } 4606, (1999). 

\bibitem{Feldman95} J.~L. Feldman {\it et al.}, {\it Phys. Rev. Lett.} {\bf
    74} 1379 (1995). 

\bibitem{Goncharov96} A.~F. Goncharov {\it et al.}, {\it Phys. Rev. B} {\bf
    54}, R15590 (1996). 

\bibitem{Mazin97} I.~I. Mazin {\it et al.}, {\it Phys. Rev. Lett.}  {\bf 78 }
  1066, (1997). 

\bibitem{Goncharov01} A.~F. Goncharov {\it et al.}, {\it Phys. Rev. B}  {\bf 63 }  064304, (2001). 

\bibitem{Harris85} A.~B. Harris and H. Meyer {\it Can. J. Phys.} {\bf 63 } 3,
  (1985). 

\bibitem{Sullivan87} N.~S. Sullivan {\it et al.}, {\it Can. J. Phys.} {\bf 65
  } 1463, (1987). 

\bibitem{Hochli90} U.~T. H\"ochli {\it et al.}, {\it Adv. Phys.}  {\bf 39 }
  405, (1990). 

\bibitem{Kokshenev96a} V.~B. Kokshenev, {\it Phys. Rev. B} {\bf 54} 1 (1996). 

\bibitem{Binder02} K. Binder, {\it J. Non-Cryst. Solids} {\bf 307} 1, (2002). 

\bibitem{Strzhemechny00} M.~A. Strezhemechny and R.~J. Hemley, {\it Phys. 
    Rev. Lett.} {\bf 85} 5595 (2000). 

\bibitem{Kaxiras94} E. Kaxiras and Z. Guo, {\it Phys. Rev. B} {\bf 49} 11822
  (1994). 

\bibitem{Cui97} T. Cui {\it et al.}, {\it Phys.  Rev. B} {\bf 55} 12253
  (1997). 

\bibitem{Runge92} K.~J. Runge {\it et al.}, {\it Phys. Rev. Lett.} {\bf 69}
  3527 (1992). 

\bibitem{Pollock94} E.~L. Pollock, and K.~J. Runge, {\it Physica B} {\bf 197}
  180 (1994). 

\bibitem{HetenyiUP} B. Het\'enyi {\it et al.}, (unpublished results). 
















\end{thebibliography}
\end{document}